# Numerical analysis of the entry flow for continuous shear thickening fluids in circular microtubes

M. Montenegro[a,c], F.J. Galindo-Rosales[b,c*]

[a]*Centro de Estudos de Fenómenos de Transporte (CEFT), Dept. Engenharia Mecânica, Faculdade de Engenharia da Universidade do Porto, Rua Dr. Roberto Frias, 4200-465 Porto, Portugal*
[b]*Centro de Estudos de Fenómenos de Transporte (CEFT), Dept. Engenharia Química, Faculdade de Engenharia da Universidade do Porto, Rua Dr. Roberto Frias, 4200-465 Porto, Portugal*
[c]*ALiCE – Laboratório Associado em Engenharia Química, Faculdade de Engenharia da Universidade do Porto, Rua Dr. Roberto Frias, 4200-465 Porto, Portugal*
\* Corresponding author: galindo@fe.up.pt

A R T I C L E   I N F O

*Keywords:*

Entry flow

Continuous Shear Thickening Fluids

CFD

A B S T R A C T

Due to their nature, using shear thickening fluids (STFs) in engineering applications has sparked an interest in developing energy-dissipating systems, such as damping devices or shock absorbers. The Rheinforce technology allows the design of customized energy dissipative composites by embedding microfluidic channels filled with STFs in a scaffold material. One of the reasons for using microfluidic channels is that their shape can be numerically optimized to control pressure drop (also known as rectifiers); thus, by controlling the pressure drop, it is possible to control the energy dissipated by the viscous effect. Upon impact, the fluid is forced to flow through the microchannel, experiencing the typical entry flow until it reaches the fully developed flow. It is well-known for Newtonian fluid that the entrance flow is responsible for a non-negligible percentage of the total pressure drop in the fluid; therefore, an analysis of the fluid flow at the entry region for STFs is of paramount importance for an accurate design of the Rheinforce composites. This analysis has been numerically performed before for shear thickening fluids modelled by a power-law model; however, as this constitutive model represents a continuously growing viscosity between end-viscosity plateau values, it is not representative of the characteristic viscosity curve of shear thickening fluids, which typically exhibit a three-regions shape (thinning-thickening-thinning). For the first time, the influence of these three regions on the entry flow on an axisymmetric pipe is analysed. 2D numerical simulations have been performed for four STFs consisting of four dispersions of fumed silica nanoparticles in polypropylene glycol varying concentration (7.5-20wt%) modelled by Khandavalli*, et al.* [1].

## 1. Introduction

Shear thickening fluids (STFs) are complex fluids, typically consisting of dense suspensions of solid particles dispersed in an inert carrier fluid, that exhibit an increase in viscosity under the application of a shear rate/stress over a critical value [2]. In other words, the stress required to shear a STF increases faster than linearly with the shear rate [3]. It is not an expected behaviour in pure substances. Still, it can be observed in concentrated suspensions where the particles show no mutual attraction towards one another under no shearing flow [4]. Depending on the shape and size of particles, their concentration, the carrier fluid etc, two kind of shear thickening behaviours have been described in the literature, i.e. Continuous Shear Thickening (CST) and Discontinuous Shear Thickening (DST). In CST, the viscosity curve typically shows three regions, a first shear thinning followed by a shear thickening and, finally, a second shear thinning, which result from the shear induced rearrangement of the particles [5]: at low shear rates the particles may form layers and the viscosity reduces from the rest state (first shear thinning region); however, for shear rates over a critical value, the hydrodynamics forces disrupt this ordered state and randomly form hydrodynamics clusters of particles, resulting in an increase of the viscosity (shear thickening region) until reaching a maximum; if the shear rate is further increased, the hydroclusters become unstable, the shear forces break them down and a new ordered state is reached providing lower viscosity values (second shear thinning region) [6]. In DST, the viscosity curve exhibits a shear thickening region much steeper than in CST and no second shear thinning region; this is a consequence of the friction and jamming between particles [3].

Whereas CST has exhibited reversibility for stable dense colloidal suspensions [7], the complete and spontaneous relaxation of the DST does not occur due to the partial retention of the frictional force chains [8, 9]. For that reason, from the practical point of view, shear thickening fluids exhibiting continuous shear thickening with no hysteresis in the flow curve measurements are more interesting for the development of composites embedding shear thickening fluids [2]. Among these applications, we are particularly interested in Rheinforce technology, formerly known as CorkSTFluidics [10], which has been successfully applied in shin guards [11] and helmet liners [12]. This technology [13] consists of adding shear thickening fluids to any resilient scaffold material by means of embedding microfluidic patterns; the right combination of the cushioning properties of the solid material, the energy dissipating properties of the shear thickening fluid and the fluid-structure interaction results in a composite material that can be tuned to damp, potentially, any impact load. The use of microfluidic network for embedding the STF into the resilient solid material introduces several major advantages with regards to other strategies, such as impregnation of fabrics or open-cell foams: 1) reduced amount of fluid [14], which is of paramount importance for applications in which lightweight is crucial; 2) enhanced rheological response of the STFs [15]; 3) the geometry of the microchannels can be numerically optimized to produce the desired fluid flow [16, 17]. This latter feature is the key feature of this technology, as it has been proven to be possible of controlling very efficiently the pressure drop by means of designing the shape of the microchannels, also known as microfluidic rectifiers [18-20]. As pressure drop is equivalent to the energy dissipation by unit volume of fluid, controlling the pressure drop in a microfluidic device is equal to controlling the amount of energy dissipated for a fixed amount of liquid. Upon the impact of the strike at a velocity $v_{strike}$, the fluid contained in the microfluidic channel under the impact zone will be forced to flow out of that region towards the both sides of the microchannel with a velocity of $v_{in} = \frac{1}{2} \cdot v_{strike} \cdot \frac{A_{strike}}{A_{in}}$ [21] (Figure 1), which resembles the entry flow in a pipe.

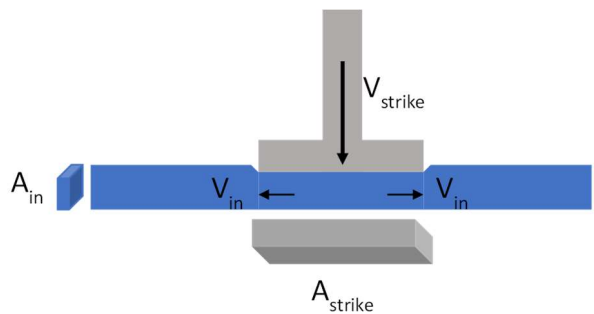

*Figure 1 – Sketch of the relationship between the velocity of the strike and the velocity of the fluid inside the microchannel.*

The pressure drop experienced by the fluid multiplied by the flow rate determines the power of energy dissipated by the fluid due to viscous effects: $W = 2 \cdot \Delta p \cdot A_{in} \cdot v_{in}$. It is well known that the pressure drop experienced by a fluid along a horizontal pipe is the sum of the pressure drop at the entrance region and the pressure drop in the fully developed region [22]. Therefore, understanding the flow dynamics of shear thickening fluids in the entrance region is paramount for better designing the Rheinforce composites.

The study of the entry flow in a pipe or a channel has been widely reported in the literature, both for Newtonian flows [23, 24], as well as for viscoelastic [25] and inelastic non-Newtonian flows, such as in the works of Gupta [26], Chebbi [27], Poole and Ridley [28], Fernandes, *et al.* [29] and Lambride, *et* *al.* [30], where a power-law model fluid (Equation (1)) allows to consider shear thinning ($n < 1$), Newtonian ($n = 1$) and shear thickening behaviours ($n > 1$) behaviours, just by playing with the values of the exponent parameter $n$:

$$\eta = k \cdot \dot{\gamma}^{n-1} \tag{1}$$

where $\eta$ is the fluid's viscosity, $k$ is the flow consistency index, $n$ is the flow behaviour index, and $\dot{\gamma}$ is the shear rate, which is defined as the second invariant of the rate of deformation tensor [31]. In these studies, the velocity profiles at different axial locations are reported and analysed and the influence of the power-law index on the development length of the velocity profiles in a 2D axisymmetric pipe is evaluated. In the work of Poole and Ridley [28], it was also reported the evolution of the velocity profiles along the development length region of the pipe when the fluid changes its rheological behaviour: from shear thinning to Newtonian and to shear thickening. The power-law model (Equation (1)) is a monotonous function that either increases or decreases the viscosity values between two limiting plateaus at very low and very high viscosity values. Consequently, it is not able to predict the three characteristic regions typical of the CST behaviour: Region-I) shear thinning at low shear rates, below the critical shear viscosity ($\dot{\gamma}_{min}$); Region-II) shear thickening at intermediate shear rates ($\dot{\gamma}_{min} \leq \dot{\gamma} \leq \dot{\gamma}_{max}$); and Region-III) characterized by another shear thinning [32]. To the best of the authors' knowledge, the effects of these three regions on the dynamics of the flow in the entry region were not analysed yet in the literature and the results obtained will provide helpful information for better designing and modelling energy dissipative composites with shear thickening fluids showing CST behaviour.

## 2. Modelling and simulation

### 2.1. Governing equations

The developed model consists of a 2D axisymmetric, incompressible, steady state, laminar flow inside a microtube. The conservation equations calculated during the numerical procedure include the Continuity Equation (2) and the Momentum Conservation Equation (3) in both the axial ($z$) and radial ($r$) directions:

$$\nabla \cdot (\rho \vec{v}) = 0 \tag{2}$$

$$\nabla \cdot (\rho \vec{v} \vec{v}) = -\nabla p + \nabla \cdot (\bar{\bar{\tau}}) \tag{3}$$

where $\rho$ is the fluid's density, $\vec{v}$ is the velocity, $p$ is the static pressure and $\bar{\bar{\tau}}$ is the stress tensor which is given, in the case of a GNF constitutive model, by Equation (4):

$$\bar{\bar{\tau}} = \eta(\dot{\gamma})[(\nabla \vec{v} + \nabla \vec{v}^T)] \tag{4}$$

being $\eta(\dot{\gamma})$ a function showing the dependence of the dynamic viscosity with the shear rate. In this study, it was considered the GNF model proposed by Khandavalli, *et al.* [1] (Equation (5)).

$$\eta(\dot{\gamma}) = \left\{\frac{\eta_\infty}{A} + \left(\eta_0 - \frac{\eta_\infty}{A}\right) \cdot [1 + (\lambda_1 \cdot \dot{\gamma})^{B_1}]^{\frac{n_1-1}{B_1}}\right\} \cdot \left\{1 + A - A \cdot \left[1 + [(\lambda_2 \cdot \dot{\gamma})^{B_2}]^{\frac{n_2-1}{B_2}}\right]\right\} \quad (5)$$

where $\eta_0$ is the zero-shear viscosity, $\eta_\infty$ represents the infinite-shear viscosity, $A$ is a parameter that determines the extent of shear-thickening, $\lambda_i$ are time constants and, therefore the inverse of the critical shear rates for thinning and thickening, $n_i$ represent the power-law exponents and $B_i$ are dimensionless transition parameters [1]. Equation (5) is a suitable GNF model for CST, as it is a smooth function of the viscosity that avoids discontinuities near the minimum and maximum values, unlike the piece-wise models proposed by Galindo-Rosales, *et al.* [32, 33]. Additionally, it has fewer parameters to fit the experimental data, which makes it more convenient for numerical simulations. The values of all these parameters for the four test fluids used in the study are presented in Table 1.

*Table 1 – Fitting parameters of shear thickening fluid model given in Equation (5) for several nanoparticle concentrations. Reprinted from [1], with permission from Elsevier..*

| Silica [wt%] | $\eta_0$ [Pa·s] | $\eta_\infty$ [Pa·s] | $B_1$ | $n_1$ | $\lambda_1$ [ms] | $B_2$ | $n_2$ | $\lambda_2$ [ms] | $A$ |
|---|---|---|---|---|---|---|---|---|---|
| 7.5 | 3.0 | 1.0 | 0.5 | 0.01 | 100 | 3.3 | -2.0 | 62 | 16 |
| 10 | 3.0 | 1.0 | 0.6 | 0.11 | 100 | 10 | -3.0 | 180 | 48 |
| 15 | 3.3 | 1.0 | 1.0 | 0.15 | 300 | 11 | -0.9 | 200 | 300 |
| 20 | 8.5 | 1.0 | 1.0 | 0.15 | 300 | 11 | -0.9 | 300 | 260 |

Figure 2 presents the viscosity curves obtained with Equation (5) for each STF formulation, where the shear thickening region takes place between the shear rates of $1 < \dot{\gamma} < 100$ s$^{-1}$. It can also be observed that the critical shear stress ($\tau_{min}$), was the same for all of them except for the most concentrated one (STF 4), probably because the dispersion quality was different from the others.

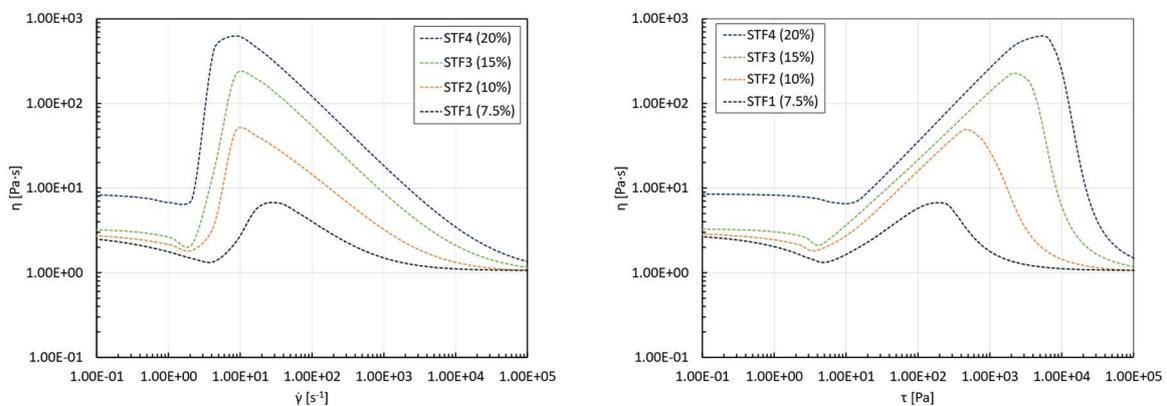

*Figure 2 – Steady shear viscosity curves for the four formulations as a function of shear rate (left) and stress (right) [1].*

## 2.2. Geometry, boundary conditions, mesh analysis and numerical methods

The fluid domain consisted of a cylinder of length $L$=1mm and a diameter $D$. Three different diameters were considered, i.e. 0.05, 0.1 and 0.2mm. The axisymmetric nature of the flow allowed to perform 2D numerical simulations.

Non-slip condition, $v_z(r = R) = 0$ was imposed on the pipe wall. A constant velocity profile is imposed at the inlet, $v_z(r) = v_{in}$. To cover the shear thickening region of the viscosity curve, the inlet velocity values were chosen between 0.1mm/s and 1.5mm/s. At the outlet, the fluid is discharged to atmospheric pressure, $p_{out} = p_{atm}$.

A size bias was used along the walls and at the pipe inlet for the mesh used in the present work. This is the region where the flow profile will develop and, therefore, the region of greater interest, hence the decision to refine this area of the tube. The convergence criteria adopted established that all residuals were below $1 \times 10^{-9}$ and that at least 2000 iterations had been calculated. To guarantee the independence of the results from the mesh, a convergence study was carried out with 3 different meshes. The properties of each mesh are shown in Table 2. All the meshes present an excellent cell quality regarding the skewness and the orthogonal quality, and the threshold aspect ratio of 1:5 is fulfilled [32].

*Table 2 – Mesh characteristics.*

| Mesh | L divisions | R divisions | Number of elements |
|---|---|---|---|
| 1 | 1000 | 50 | 50.000 |
| 2 | 1000 | 100 | 100.000 |
| 3 | 2000 | 100 | 200.000 |

The convergence study was carried out using the fluid with the steepest grow of viscosity in the shear thickening region (STF 4) and an inlet velocity of 0.25mm/s. Figure 3 illustrates the evolution of the normalised axial velocity $\frac{v_c}{v_{in}}$ along the tube; the inset graph shows the velocity profile at the normalised coordinate $\frac{z}{D}$=0.1 and the fully developed profile at the outlet, obtained for the 3 meshes.

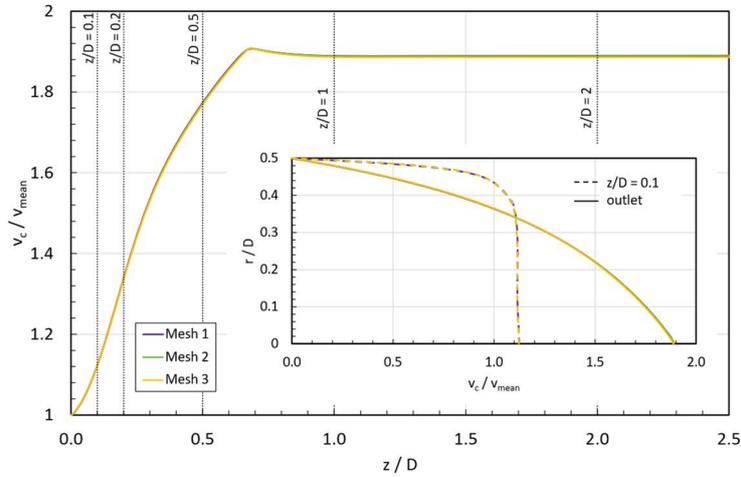

*Figure 3 – Mesh comparison - STF 4 - $v_{in} = 0.25$mm/s.*

The velocity distribution and profiles along the pipe are consistent across the three meshes, ensuring the mesh independence of the numerical simulation. Mesh 1 is chosen for the subsequent simulations because it balances computational efficiency and accuracy, especially near the critical zones (boundary wall and inlet region) where a finer mesh is needed.

The numerical simulations were performed using the commercial software FLUENT distributed by *ANSYS®* using the available flow models and simulating the shear thickening effect by implementing the constitutive equation (Equation (5)) for the STFs through a user-defined function (UDF). For the numerical discretization of the governing equations presented in Section 2.1, the Green-Gauss Node-Based gradient evaluation is used. This discretization method computes the face value of a given variable through the arithmetic average of the nodal values on the face; this gradient scheme is more accurate than the cell-based gradient on irregular (skewed and distorted) meshes [34]. The coupling between the pressure and velocity fields is achieved through the segregated SIMPLE (Semi-Implicit Method for Pressure-Linked Equations) algorithm.

## 3. Results and discussion

Figure 4 shows the normalised axial velocity, $\frac{v_c}{\bar{v}}$, along the normalised pipe length, $\frac{z}{D}$, for the four STF formulations. The presented velocity values are the maximum ones, recorded in the central axis of the pipe. The results correspond to the inlet velocities simulations described in Section 2.2, in a pipe with the geometry $L \times D = 1 \times 0.1$ mm. In each subfigure it was also added the curve corresponding for a Newtonian fluid.

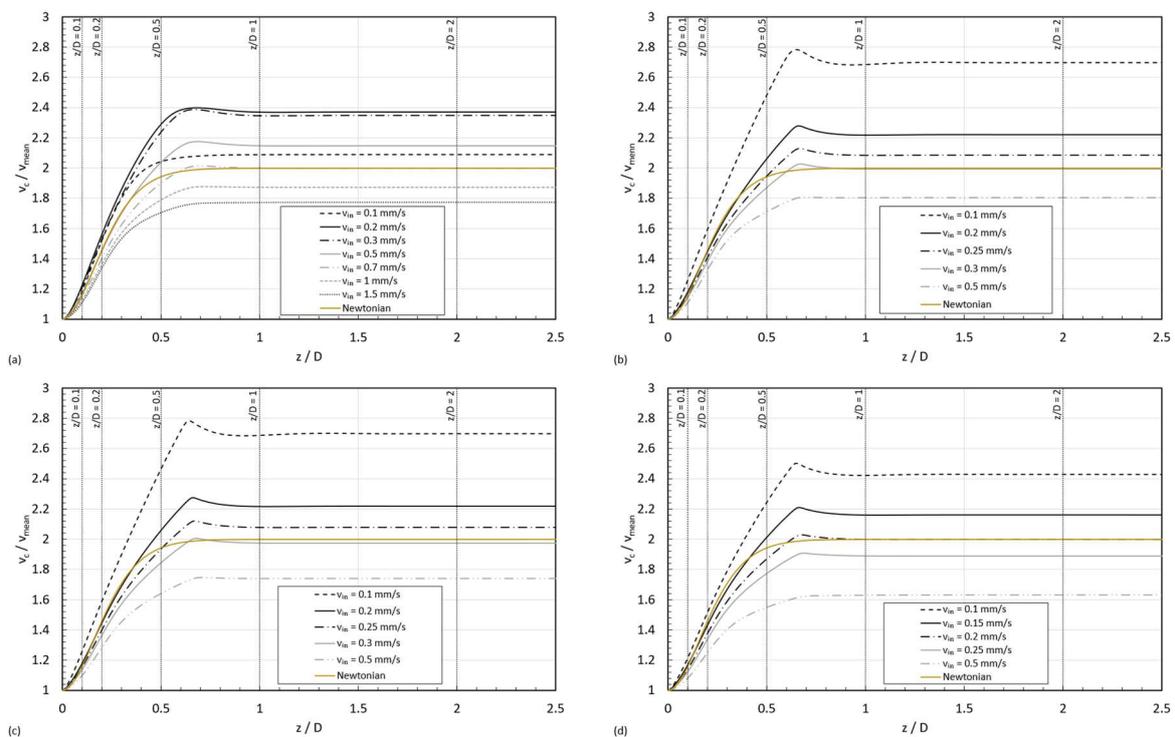

*Figure 4 – Normalised axial velocity for the four formulations of four shear thickening fluids with increasing concentration for increasing inlet velocities (a) STF 1, (b) STF 2, (c) STF 3 and (d) STF 4.*

The STFs have different normalized axial velocity curves from the Newtonian fluid, which has an unchanging curve for any inlet velocity. This difference is due to the shear thickening effect, which causes a local microstructure to form under shear. The inlet velocity affects how the STFs behave, changing the shear rate and the microstructure formation. Thus, for the same channel size and the same inlet flow rates, the normalized axial velocity curves vary depending on the viscosity curve of the STF; however, a general trend can be observed: it grows with the normalized position up to a maximum value, and then it slightly decreases until reaching the fully developed region. This phenomenon challenges the conventional definition of entry region for these fluids.

The entrance length, $L_e$, is the distance from where the fluid enters the pipe to where the boundary layer reaches the centreline and the velocity profile becomes fully developed and constant [35]. Lambride and colleagues [30] conducted a numerical study on the flow behaviour of power-law fluids in pipes and channels. They computed the entrance length as a function of the transverse coordinate using a different definition based on the wall shear stress evolution. They found that the stress entrance length was lower than the conventional centreline entrance length for shear thinning fluids (especially at low Reynolds numbers). However, for pipe flow, they showed that the usual definition of the development length was a good measure of flow development for power-law exponent values above 0.7, regardless of the Reynolds number. This indicated that the flow developed more slowly at the symmetry axis in shear thickening fluids [30]. Based on their results, in this study, we stick to the conventional centreline entrance length definition.

For Newtonian [23, 24], viscoelastic [25] and inelastic non-Newtonian shear thinning fluids [26-30], the normalized axial velocity increases monotonically with the normalized z-position, so it is usually assumed that the velocity profile is fully developed when the centreline velocity, $v_c$, is at least 99% of the maximum velocity, $v_{max}$, which is equal to the exit velocity, $v_{out}$, for Newtonian and shear thinning fluid flows. However, in most simulations, STFs show a peak in the centreline velocity before the profile is fully developed so the criterion for estimating the entrance length in this work must account for this feature. Therefore, a different method was used to solve this problem: starting from the outlet and moving towards the inlet, if there is no peak, the entrance length is where the centreline velocity is lower than 99.9% of $v_{out}$; if there is a peak in the centreline velocity, the entrance length is where the centreline velocity is higher than 100.1% of $v_{out}$.

In Creeping flow ($Re \ll 1$), the vast majority of research works reported that the entry length for Newtonian fluids follows Equation (6):

$$\frac{L_e}{D} = C_1 + C_2 \cdot Re \qquad (6)$$

where $C_1$ is the asymptotic limit of the entrance length value when $Re \to 0$ [24]. For better understanding of the values available in the literature for the coefficients in Equation (6) the reading of the works of Poole and Ridley [28], Li, et al. [36] and Ferreira, et al. [24] are suggested. Regarding non-Newtonian fluids, the definition of the Reynolds number is not straightforward, due to the dependence of the viscosity with the shear rate. The use of the Reynolds number developed by Metzner and Reed [37] ($Re_{MR} = \frac{\rho \cdot v_{mean}^{2-n} \cdot D^n}{k} \cdot 8 \cdot (\frac{n}{6 \cdot n+2})^n$, being $v_{mean} = v_{in}$ in this study) allows the development length at high Reynolds number to collapse onto a single curve (Equation (6)) in which the coefficient are independent of the $n$ index [28]. Since the model developed by Khandavalli, et al. [1] consists of a product of two Carreau-Yasuda models (one each for shear-thinning and shear-thickening regions, indicated by the subscripts 1 and 2, respectively), the shear thickening region of each fluid was approximated to a power trendline (Figure 5) in order to obtain both the flow consistency index $k$ and the flow behaviour index $n$ for each formulation that allow the calculation of the aforementioned power-law Reynolds number.

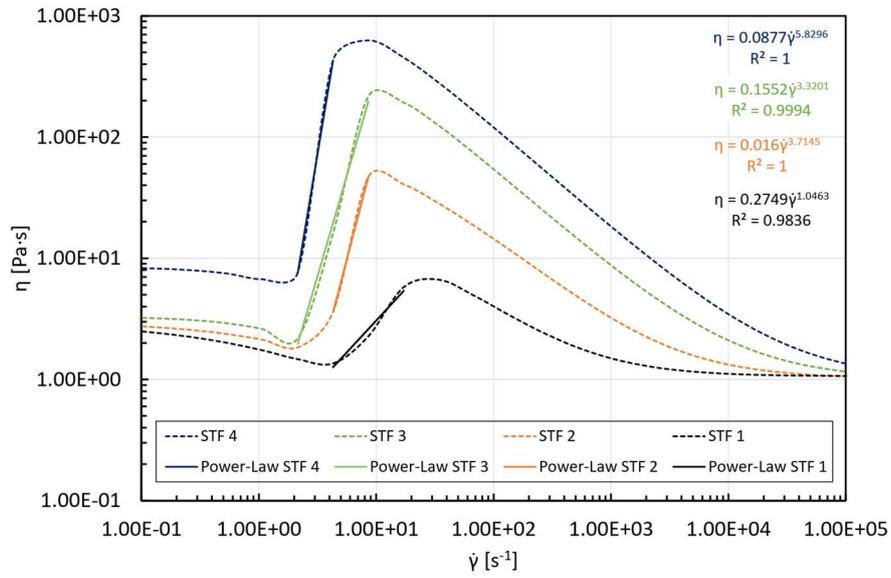

*Figure 5 – Shear thickening region adjusted to the power-law model.*

Figure 6 presents the normalised entrance lengths, $\frac{L_e}{D}$, as a function of $Re_{MR}$; it can be observed that the Reynolds numbers lays below $10^{-2}$, evidencing that the fluid is working in creeping flow and consequently, at such a low Reynolds regime, the normalised entry length remains constant and independent of the $n$ index. This result is consistent with the plots shown in Figure 4, as the axial centreline velocity profile seems to be constant for $\frac{z}{D}$>0.91.

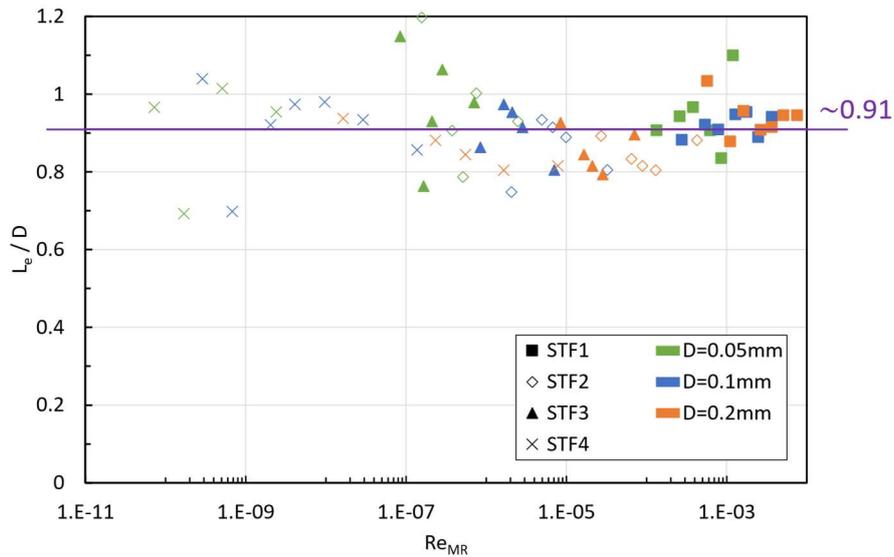

*Figure 6 – Entrance length vs Reynolds number.*

As mentioned above, within the entry region ($z < L_e$), depending on the inlet velocity and the rheological properties of the STF, the normalised axial velocity curves grows from 1 up to a maximum that can be larger than the value corresponding to the fully developed velocity. Poole and Ridley [28] observed a similar overshoot in the centreline velocity for power-law shear thickening fluids ($n > 1$). However, the cause of this overshoot has never been analysed in detail. Figures 7.a) and 7.b) show the changes in the viscosity and stresses within the entry region of the STF 4 formulation for the case in which the inlet velocity is 0.1mm/s. It is noticeable that the fluid adapts to the flow with a V-shaped peak appearing in the entry region, providing a minimum value of viscosity in every location at which the stresses are minimum, i.e. the critical shear stress ($\tau_{min}$) providing the onset of the shear thickening behaviour (Figure 2). That viscosity and normalised shear stress ($\frac{\tau}{\tau_{min}}$) contour plots are a consequence of imposing a complex flow to the shear thickening fluids, as in the centreline ($r = 0$) the fluid is being undergone a pure elongational flow with $\dot{\varepsilon} = \frac{dv_z}{dz}$ and at the wall ($r = \frac{D}{2}$) the flow was simple shear with $\dot{\gamma} = \frac{dv_r}{dz}$. The fluid domain in between the centreline and the wall of the pipe will be undergone a complex flow. The complexity of the flow is well represented by the flow-type parameter [38], defined by Equation (7):

$$\xi = \frac{\|\boldsymbol{D}\| - \|\boldsymbol{\Omega}\|}{\|\boldsymbol{D}\| + \|\boldsymbol{\Omega}\|} \quad (7)$$

where $\|\boldsymbol{D}\|$ is the magnitude of the rate-of-deformation tensor and $\|\boldsymbol{\Omega}\|$ is the magnitude of the vorticity tensor. Thus, when $\xi = 1$ the region of the fluid domain is dominated by purely elongational flow, if $\xi = 0$ the flow is dominated by simple shear; if $\xi = -1$ the flow approaches a solid-body rotation; finally, other portions showing a combination of these [39]. Figure 7.c) shows that, within the entry region, the viscosity is minimum when the flow approaches a solid-body rotation and it is maximum in those locations where the shear or extensional flows are strong. The peak in the velocity profile appears when the local shear rate provides the minimum local viscosity, in the surroundings of the centreline at the end of the entry length, where the solid rotation encounters the centreline, thus, the local viscosity is minimum and the fluid is squeezed axially providing the overshoot in the axial velocity at the centreline.

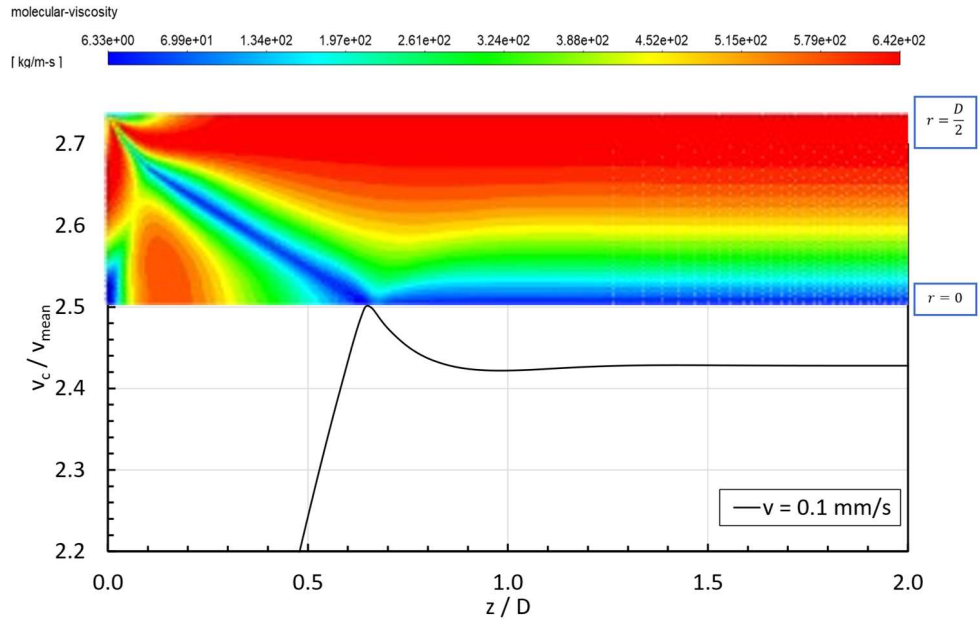

*Figure 7.a) – STF 4 viscosity along the pipe for $v_{in} = 0.1$ mm/s.*

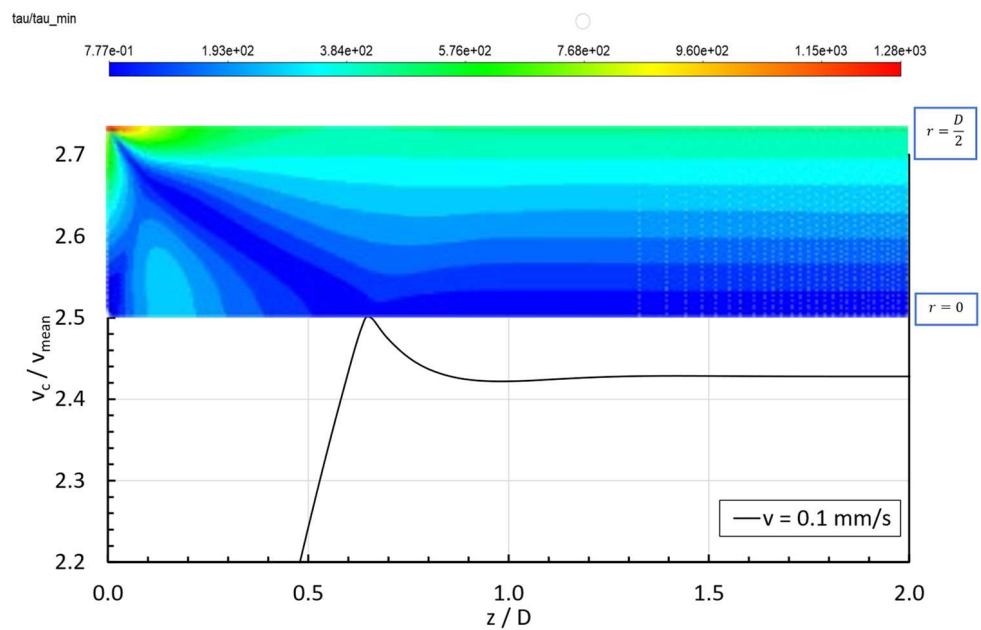

*Figure 7.b) – STF 4 normalised shear stress along the pipe for $v_{in} = 0.1$ mm/s.*

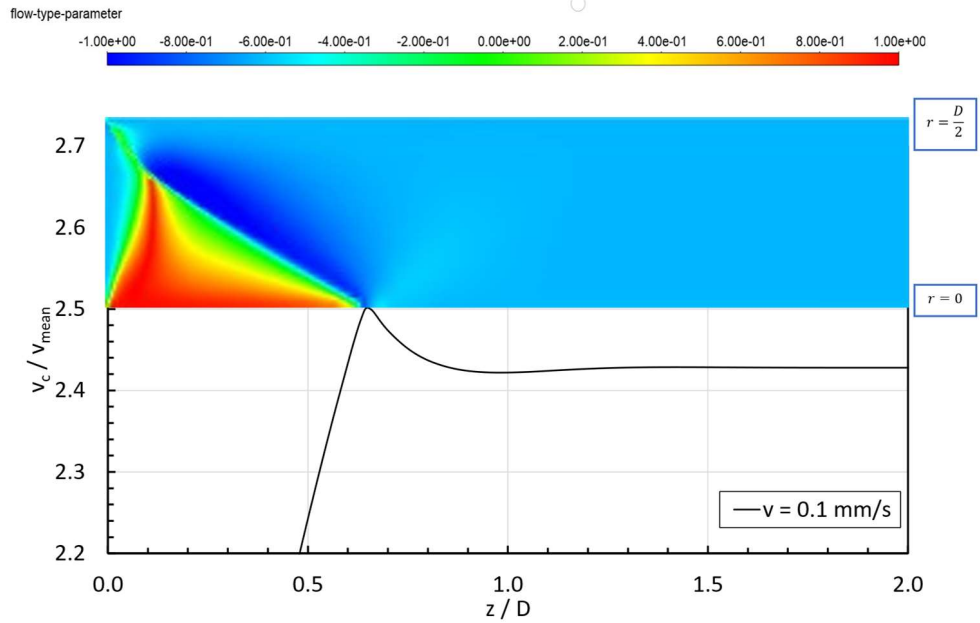

*Figure 7.c) – STF 4 flow type parameter along the pipe for $v_{in} = 0.1$ mm/s.*

For the fully developed region ($z > L_e$), when the velocity profile does not depend on the z-position, the extensional flow disappears ($\dot{\varepsilon} = \frac{dv_z}{dz} = 0$) and the fluid is undergone simple shear flow throughout the whole fluid domain [40].

Further analysing Figure 4, it can be observed that the peak value in the normalised axial velocity at the centreline also depended on the imposed inlet velocity and the viscosity curve of the fluid. Figure 8 shows that the milder shear thickening fluid (STF 1) exhibited a maximum in $\frac{v_c}{v_{mean}}$ increasing with $v_{in}$ until reaching a maximum and then it started decreasing below the Newtonian limit ($\frac{v_c}{v_{mean}} = 2$). STF 2, STF 3 and STF 4 exhibited similar trends in the peak dependency with $v_{in}$, starting for the highest value in $\frac{v_c}{v_{mean}}$ and, subsequently, decreasing exponentially with $v_{in}$ again below the Newtonian limit. However, for STF 4, the fluid with the strongest shear thickening response, the starting peak value was lower than the one for STF 2 and STF 3.

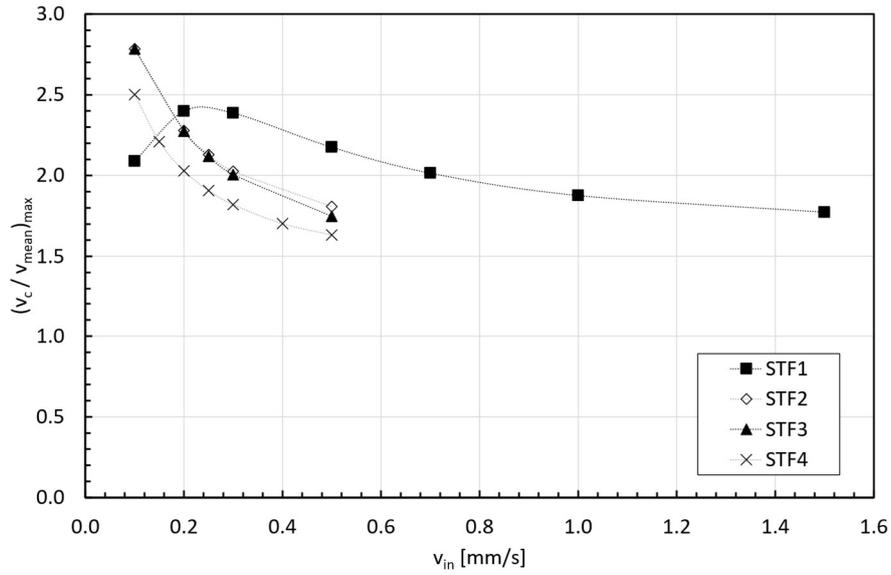

*Figure 8 – Evolution of peak normalised velocity with inlet velocity.*

This effect was not observed before for STFs, because the considered power-law equation provides a viscosity that increases monotonically for $n > 1$ [28]; however the viscosity curves are not, exhibiting three regions (Shear thinning [ Ⅰ ] – Shear thickening [ Ⅱ ] – Shear thinning [Ⅲ]) [32, 33], as depicted in Figure 2. It is, therefore, paramount to analyse the shape of the velocity profiles in combination with the viscosity curves to fully understand the results shown in Figure 8.

Figure 9 shows the velocity profiles for the four STF formulations with an inlet velocity of 0.5mm/s at different locations in $z$-direction, from the inlet to the fully developed region, compared to the Newtonian case. The remaining velocity profiles for the additional inlet velocities are shown in the Supplementary material.

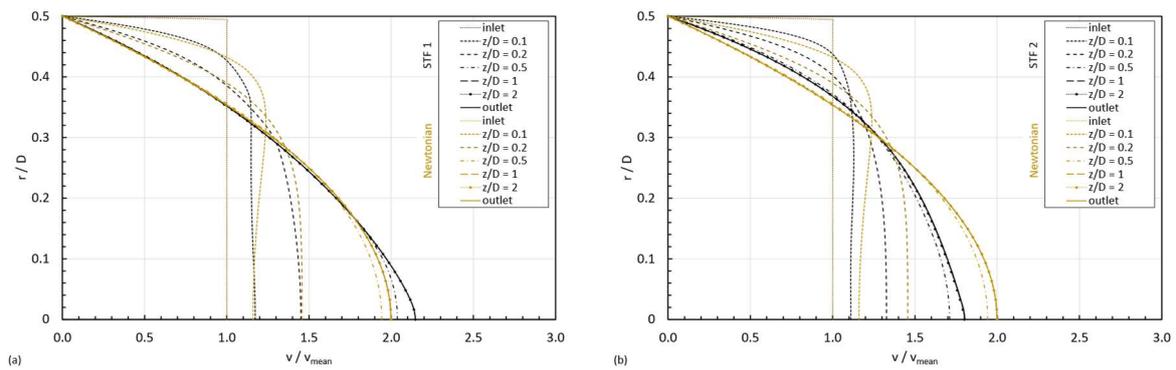

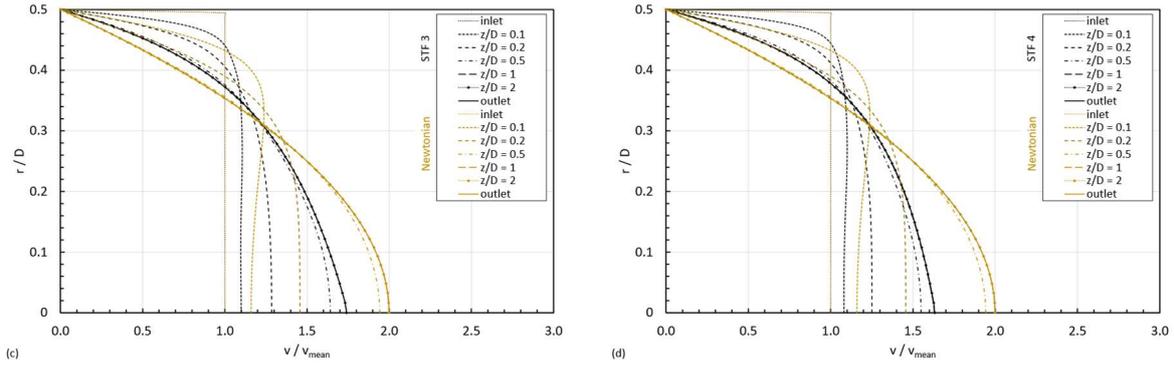

*Figure 9 – Velocity profiles for the four formulations of shear thickening fluids with increasing concentration in a D = 0.1mm pipe: (a) STF 1, (b) STF 2, (c) STF 3 and (d) STF 4.*

Figure 9 shows the velocity profiles along the pipe for the four formulations of shear thickening fluids and for an inlet velocity of 0.5mm/s. It can be observed that, for the same inlet velocity, when the shear thinning behaviour dominates next to the centreline, the maximum normalised velocity is smaller than the maximum normalised velocity developed by a Newtonian fluid. The contrary happens, when the shear thickening behaviour is triggered next to the centreline; the maximum normalised velocity for a shear thickening behaviour is larger than that for a Newtonian fluid. These graphs are consistent with the results shown by Poole and Ridley [28] and Lambride, *et al.* [30]. The shape of the developed velocity profile is responsible for a gradient of shear rates in the radial direction that grows from a minimum at the centreline ($\dot{\gamma}_c$) towards a maximum at the wall of the pipe ($\dot{\gamma}_w$). In the case of modelling the shear thickening behaviour by a power-law model, as the curve viscosity monotonically increases from low to high viscosity values, there is a monotonically increasing viscosity from low to high values, from the centreline to the wall. However, the viscosity model proposed by Khandavalli, *et al.* [1] is richer than the power-law model, in the sense that it is able to cover the three typical regions in the shear thickening behaviour, i.e., the first shear thinning (region Ⅰ), the shear thickening (region Ⅱ) and the second shear thinning (Ⅲ); consequently, depending on the formulation of the fluid, the inlet velocity and the dimensions of the pipe, different velocity profiles are developed, resulting in different values for $\dot{\gamma}_c$ and $\dot{\gamma}_w$ ($\dot{\gamma}_c < \dot{\gamma}_w$), which may lead to six different cases depending on their respective location in the viscosity curve:

- Case 1: both $\dot{\gamma}_c$ and $\dot{\gamma}_w$ are within region I. The first shear thinning behaviour is dominating the whole fluid domain, and the viscosity decreases monotonically from the centreline towards the wall in the radial direction.
- Case 2: $\dot{\gamma}_c$ belongs to region I and $\dot{\gamma}_w$ is within region II. The shear thinning behaviour is dominating next to the centreline and the shear thickening does it next to the wall; consequently, there is a non-monotonical variation of viscosities in the radial direction, and there will be a minimum in the viscosity at a certain distance from the centreline, when the shear rate reaches the $\dot{\gamma}_c$ in the viscosity curve.
- Case 3: $\dot{\gamma}_c$ belongs to region I and $\dot{\gamma}_w$ is within region III. The shear thinning behaviour is dominating next to the centreline and next to the wall; however, the fact of reaching the two critical shear rates ($\dot{\gamma}_{min}$ and $\dot{\gamma}_{max}$) in the viscosity curve results in a non-monotonical variation of viscosities in the radial direction. The viscosity will diminish from the centreline towards a minimum at a certain distance from the centreline; then it will increase until the maximum in shear rate, closer to the wall; and, finally, the viscosity will decrease from that maximum until reaching $\dot{\gamma}_w$ at the wall.

- Case 4: both $\dot{\gamma}_c$ and $\dot{\gamma}_w$ are within region II. The shear thickening behaviour is dominating the whole fluid domain, and the viscosity increases monotonically from the centreline towards the wall in the radial direction.
- Case 5: $\dot{\gamma}_c$ belongs to region II and $\dot{\gamma}_w$ is within region III. The shear thickening behaviour is dominating next to the centreline and the shear thinning does it next to the wall. Consequently, the viscosity will increase from the centreline to reach a maximum at a certain distance and, from that position, it will decrease towards the wall of the pipe.
- Case 6: both $\dot{\gamma}_c$ and $\dot{\gamma}_w$ belong to region III. This scenario is similar to the case 1, in the sense that the viscosity decreases radially from the centreline towards the wall of the pipe, but in this case it follows the second shear thinning and not the first one in the viscosity curve.

To observe which of the above discussed scenarios is taking place in any of the different flow configurations considered in this study, it is preferrable to normalize the shear rate for each formulation accordingly to Equation (8).

$$\dot{\gamma}^* = \frac{(\dot{\gamma} - \dot{\gamma}_{min})}{(\dot{\gamma}_{max} - \dot{\gamma}_{min})} \qquad (8)$$

Where $\dot{\gamma}_{min}$ is the minimum shear rate corresponding to the start of the shear thickening region and $\dot{\gamma}_{max}$ is the shear rates corresponding to maximum viscosity. The above discussion is valid only for axial positions beyond the entry length, when the velocity profile reaches a fully developed shape and the flow-type corresponds to simple shear. Within the entry region, the situation is much complex due to the gradient in the flow type from the centreline to the wall of the pipe, which is different at different locations in the axial direction.Figure 10 shows the normalised shear rate against the normalised radius for STF 1. As mentioned above, in the fully developed region, depending on the inlet velocity, three different cases can be observed, i.e. cases 1, 2 and 5. In the entry length, just for a given inlet velocity, depending on the position in the axial direction, the shear rate can be larger at the centre than at the wall ($\dot{\gamma}_c > \dot{\gamma}_w$), due to the extensional flow contribution at the centreline; consequently, the following cases arise:

- Case 7: both $\dot{\gamma}_c$ and $\dot{\gamma}_w$ are within region I. Since $\dot{\gamma}_c > \dot{\gamma}_w$, the viscosity increases from the centreline towards the wall. It is the reversed situation discussed in case 1.
- Case 8: $\dot{\gamma}_w$ belongs to region I and $\dot{\gamma}_c$ is within region II. The shear thinning behaviour is dominating next to the wall and the shear thickening does it next to the centreline, resulting in the reverse situation described in case 2.
- Case 9: $\dot{\gamma}_w$ belongs to region I and $\dot{\gamma}_c$ is within region III. The second shear thinning behaviour is dominating next to the centreline, whereas the first shear thinning does it next to the wall; resulting in the reversed case 3.
- Case 10: both $\dot{\gamma}_c$ and $\dot{\gamma}_w$ are within region II. The shear thickening behaviour is dominating the whole fluid domain, but because $\dot{\gamma}_c > \dot{\gamma}_w$, the viscosity decreases monotonically from the centreline towards the wall in the radial direction.
- Case 11: $\dot{\gamma}_w$ belongs to region II and $\dot{\gamma}_c$ is within region III. The shear thickening behaviour is dominating next to the wall and the shear thinning does it next to the centreline. Consequently, the viscosity will increase from the centreline to reach a maximum at a certain distance and, from that position, it will decrease towards the wall of the pipe.
- Case 12: both $\dot{\gamma}_c$ and $\dot{\gamma}_w$ belong to region III. This cenario is similar to the case 7, in the sense that the viscosity increases radially from the centreline towards the wall of the pipe, but in this case it follows the second shear thinning and not the first one in the viscosity curve.

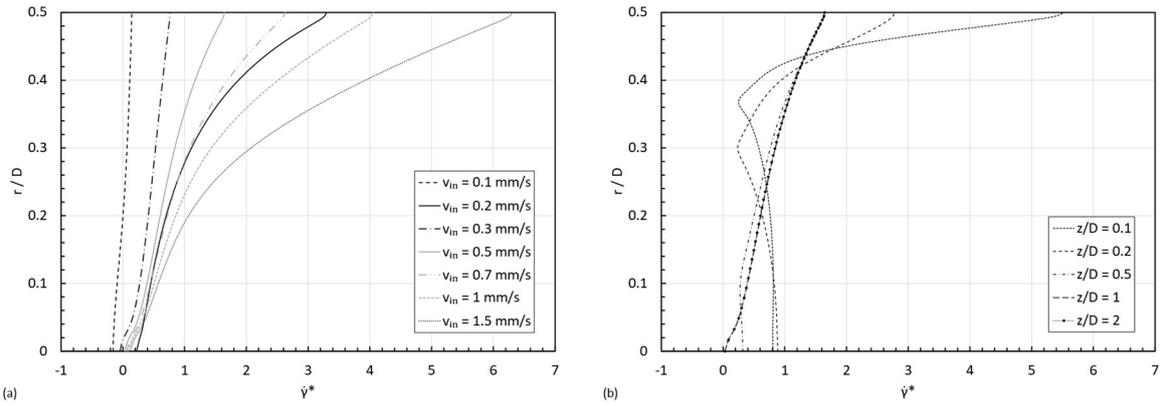

Figure 10 – STF 1 normalised shear rate profile: (a) in the fully developed region for all inlet velocities and (b) at different $\frac{z}{D}$ for $v_{in}$ = 0.5mm/s.

For the flow configurations considered in this study, we could observe the following cases 1, 2, 5, 10 and 12 as shown in Figure 10. However, if we play with the dimensions of the pipe other cases arise, like case 3 illustrated in Figure 11, where the STF 1 normalised shear rate profile in a $D$ = 0.2mm pipe for an inlet velocity of 1mm/s is shown.

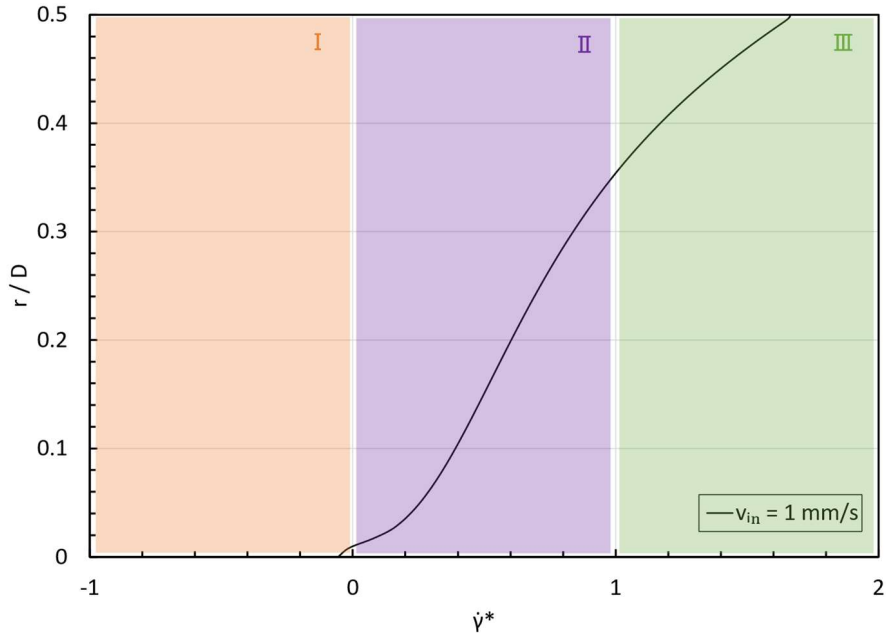

Figure 11 – STF 1 normalised shear rate profile in a $D$ = 0.2mm pipe for $v_{in}$ = 1mm/s.

As stated before, the range of inlet velocities was selected in such a way that the shear rate on the wall covered the shear thickening region of the viscosity curve of the fluid. In order to better perceive which region of the viscosity curve the flow is encompassing, the evolution of the non-dimensional viscosity, $\frac{\eta}{\eta_0}$ with the normalised shear rate along the wall is calculated by Equation (9) and presented in Figure 12.

$$\dot{\gamma}_w^* = \frac{(\dot{\gamma}_w - \dot{\gamma}_{min})}{(\dot{\gamma}_{max} - \dot{\gamma}_{min})} \qquad (9)$$

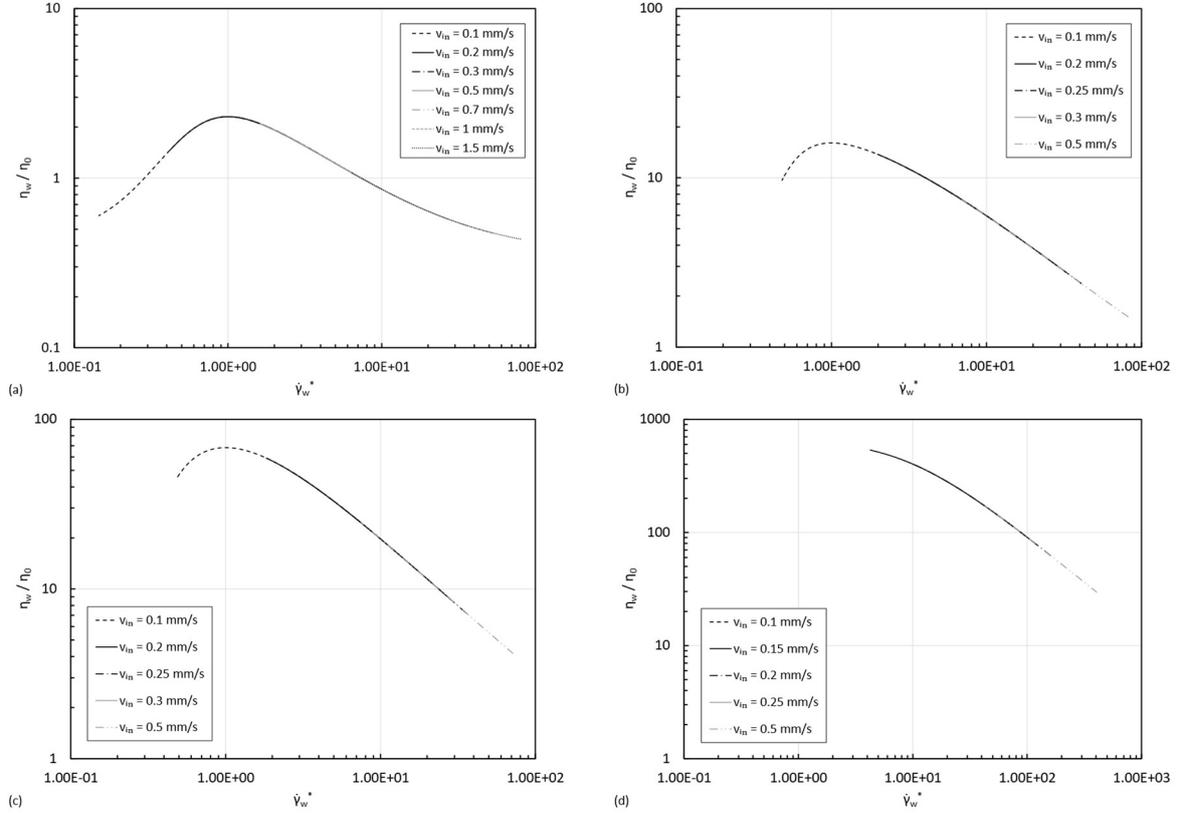

*Figure 12 – Non-dimensional viscosity versus normalised shear rate along the wall for the formulations: (a) STF 1 (b) STF 2, (c) STF 3 and (d) STF 4.*

Figure 12 allows to explain the results in Figure 8, as it evidences that only for the STF 1 formulation the selected velocities ensure that, on the pipe wall, the whole range of the shear thickening region is covered throughout the simulations, as the onset of the shear thickening behaviour is clearly visible. The inlet velocities chosen for the remaining formulations resulted in viscosity values past the critical shear rate of the viscosity curve, already reaching the maximum viscosity value and covering mainly the second shear thinning region of the curve; it is even noticeable that for the STF 4 formulation, the peak viscosity on the wall is never achieved, since it occurs for a normalised wall shear rate value of 1 and the results clearly indicate that we are in the descendent part of the viscosity curve, corresponding to the second shear thinning region.

It is well known for Laminar flow and Newtonian fluids that the flow pattern within the entry flow region is responsible for a pressure loss that depends on the entrance geometry and each geometry has an associated loss coefficient due to viscous dissipation (Equation (10)):

$$k_L = \frac{\overline{\Delta P}}{\frac{1}{2} \cdot \rho \cdot v_{in}^2} \qquad (10)$$

being $\overline{\Delta P} = \overline{P}(z = 0) - \overline{P}(z = L_e)$ the average pressure drop within the entry region. In the case of a square-edged entrance, the loss coefficient is approximately one-half of a velocity head, with fluid losing energy as it enters the pipe [22].

As discussed before, in the case of shear thickening fluids, for a given geometry, the inlet velocity determines the thickening state of the fluid and, consequently, it will affect the pressure loss within the entry region. Figure 13 illustrates the percentage of dissipated power along the tube, calculated by dividing the average of the radial pressure drop along the pipe ($\overline{\Delta P} = \overline{P}(z = 0) - \overline{P}(z)$) by the radial average of the total pressure (static pressure, $\overline{P}(z = 0)$, plus dynamic pressure, $\frac{1}{2} \cdot \rho \cdot v_{in}^2$) (Equation (11)), for the four STF formulations, as well as for a Newtonian fluid:

$$\% \frac{Dissipated\ power}{Input\ power} = \frac{\overline{P}(z = 0) - \overline{P}(z)}{\overline{P}(z = 0) + \frac{1}{2} \cdot \rho \cdot v_{in}^2} \cdot 100 \qquad (11)$$

It can be observed that the amount of dissipated power grows nonlinearly within the entry region ($z < L_e$).

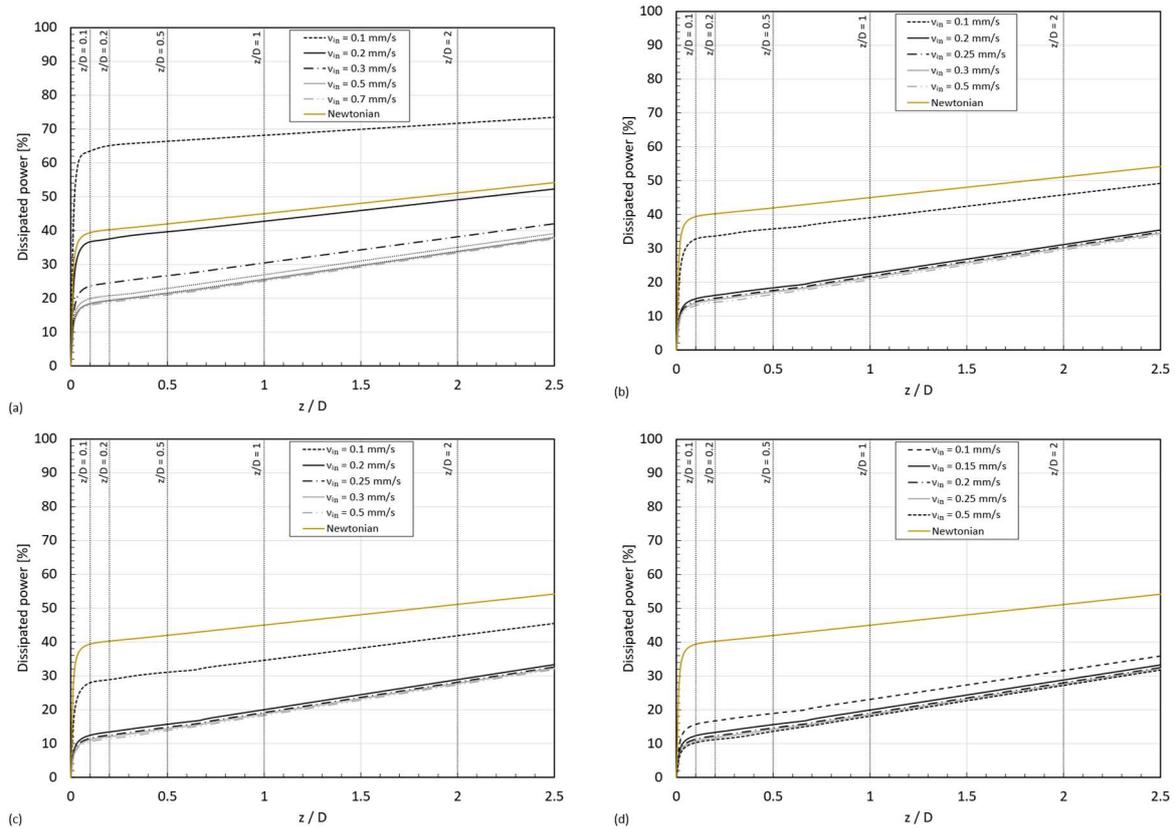

Figure 13 – Percentage of dissipated power along the tube in a D = 0.1mm pipe for the formulations: (a) STF 1; (b) STF 2; (c) STF 3; (d) STF 4.

The percentage of dissipated power at the entry length coordinate for each STF and the Newtonian case can be observed in Figure 14. Whereas for the Newtonian fluid, the value slightly grows with the inlet velocity for a given geometry, as expected, the percentage of dissipated power at the entry length for the STFs depends both on the inlet velocity and the viscosity curve of the fluid. It is all cases the percentage of energy dissipated decreases with the increase in the inlet velocity until reaching a saturating value; this result can be explained by results shown in Figure 12, where we could observe the transition towards the second shear thinning region in the viscosity curve with the increase of the inlet velocity.

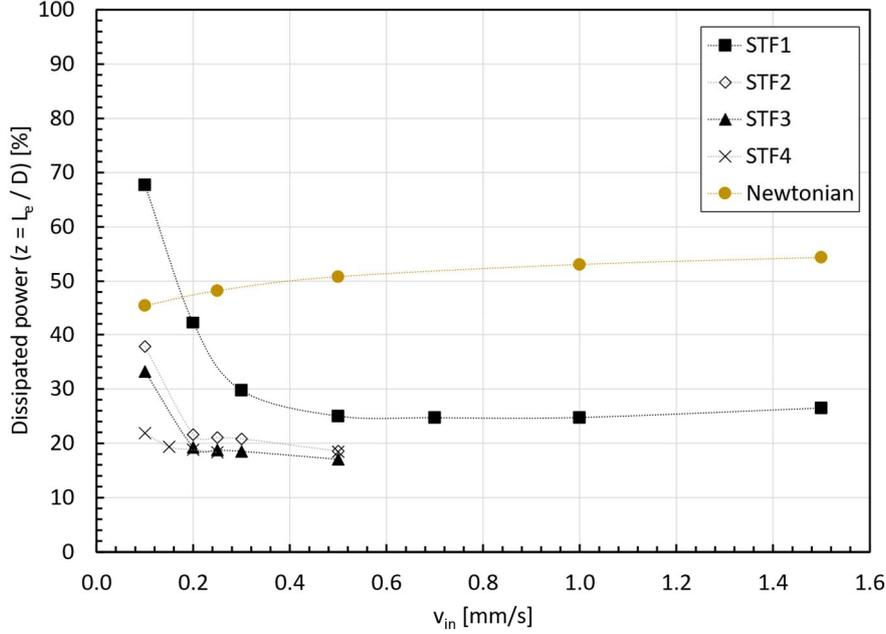

Figure 14 – Percentage of dissipated power at coordinate $z = L_e$.

## 4. Conclusions and final remarks

A systematic and detailed numerical investigation of the entry region effects in the creeping flow regime of shear thickening fluids in 2D axisymmetric microtube flow was conducted in this work. The GNF constitutive model proposed by Khandavalli, *et al.* [1] described the three regions of the viscosity curve characteristic of STFs exhibiting CST behaviour, allowing to analyse the influence of the non-monotonic shape of the viscosity curve with the shear rate on the entrance length. Due to the small characteristic length and the range of velocities of interest, the flow regime analysed laid in creeping flow regime, for very low Reynolds numbers ($Re_{MR} < 10^{-2}$). Therefore, as expected, the entry length was independent from the Reynolds number, a result that is consistent with the literature for power-law fluids [28]; however, the entry length ($\frac{L_e}{D} \sim 0.91$) was the same independently from the formulation of the shear thickening fluid considered, i.e. independent from the slope of the viscosity curve within the thickening region ($n$-index for a power-law fluid). At first sight, this result may seem unexpected and shocking, but if we analyse the stress distribution within the entry region for the four fluids considered, we can observe that there is a maximum in the axial velocity profile at the centreline occurring at the location in which the critical shear stress was minimum. That minimum shear stress value was practically the same for the four fluids, as it is expected from having the same particle, particle size, carrier fluid, similar dispersion quality, but different concentration. Moreover, the minimum in the critical shear stress took place exactly when the extensional flow disappeared. Quantitatively, the value of the peak in the velocity centreline depends on the inlet velocity imposed and the rheology of the fluid. By representing the evolution of the non-dimensional viscosity ($\frac{\eta_w}{\eta_0}$) with the normalised shear rate along the wall ($\dot{\gamma}_w^* = \frac{(\dot{\gamma}_w - \dot{\gamma}_{min})}{(\dot{\gamma}_{max} - \dot{\gamma}_{min})}$) it was possible to unveil which region of the viscosity curve was mostly activated for a given inlet velocity. Thanks to Khandavalli, *et al.* [1] model, which allowed to cover the three characteristic regions of the CST behaviour, it was possible to analyse for the very first time the richness of possible cases in the radial distribution of the viscosity, which depends on the interplay between the inlet velocity, the size of the tube, the rheological properties of the fluid and the location in *z*-direction, i.e. inside or outside the entry region. Moreover, it was also reported for the very first time that the local energy loss due to the entry flow for CST fluids depends nonlinearly on the inlet velocity and also on the viscosity curve of the fluid. This latter result is paramount for choosing the right formulation of CST fluids for each application.

Despite the novelty and practical interest of the results and conclusions withdrawn in this work, this study is however limited to shear thickening fluids exhibiting inelastic continuous shear thickening behaviours. Nevertheless, it is well known that these fluids, apart from exhibiting a non-linear relationship between stress and shear rate, also exhibit viscoelastic behaviour [41]; nevertheless, the literature lacks a constitutive model able to predict both features simultaneously. The rheological community should work to fill the empty gap in the state-of-art in order to obtain meaningful numerical results, as it has been well documented the importance of shear-induced and extension-induced elastic stresses in energy dissipation for viscoelastic fluids undergone complex flows [39] and at small length scales [42].

**Acknowledgments**

This project was financially supported by: LA/P/0045/2020 (ALiCE) and UIDB/00532/2020, UIDP/00532/2020 (CEFT) and UI/BD/150887/2021, and the program Stimulus of Scientific Employment, Individual Support-2020.03203.CEECIND funded by national funds through FCT/MCTES (PIDDAC). Authors are also grateful to Dra. L. Campo-Deaño for fruitful and selfless discussion.

# Supplementary material

*Velocity profiles*

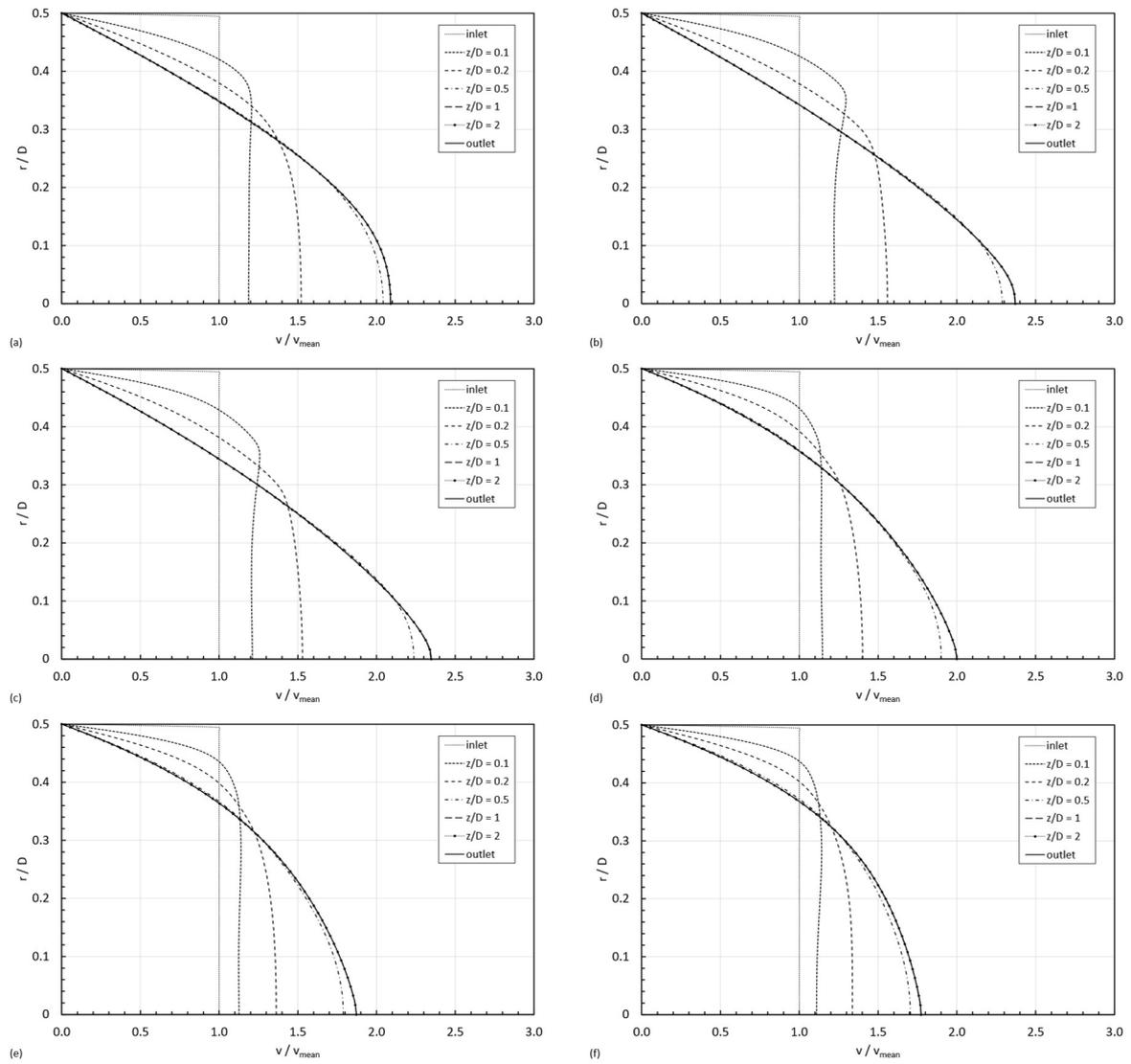

*Figure S1 – Velocity profiles for the STF 1 formulation in a D = 0.1mm pipe: $v_{in}$ = (a) 0.1mm/s, (b) 0.2mm/s, (c) 0.3mm/s, (d) 0.7mm/s, (e) 1mm/s and (f) 1.5mm/s.*

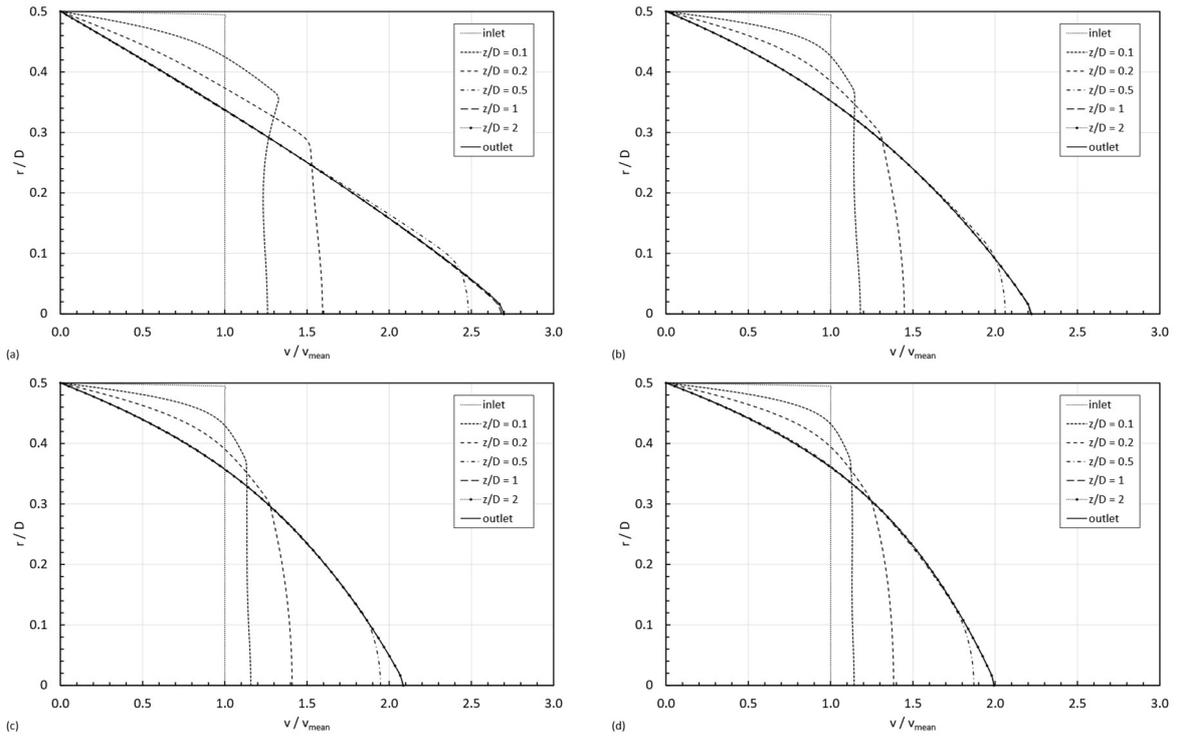

*Figure S2 – Velocity profiles for the STF 2 formulation in a D = 0.1mm pipe: $v_{in}$ = (a) 0.1mm/s, (b) 0.2mm/s, (c) 0.25mm/s and (d) 0.3mm/s.*

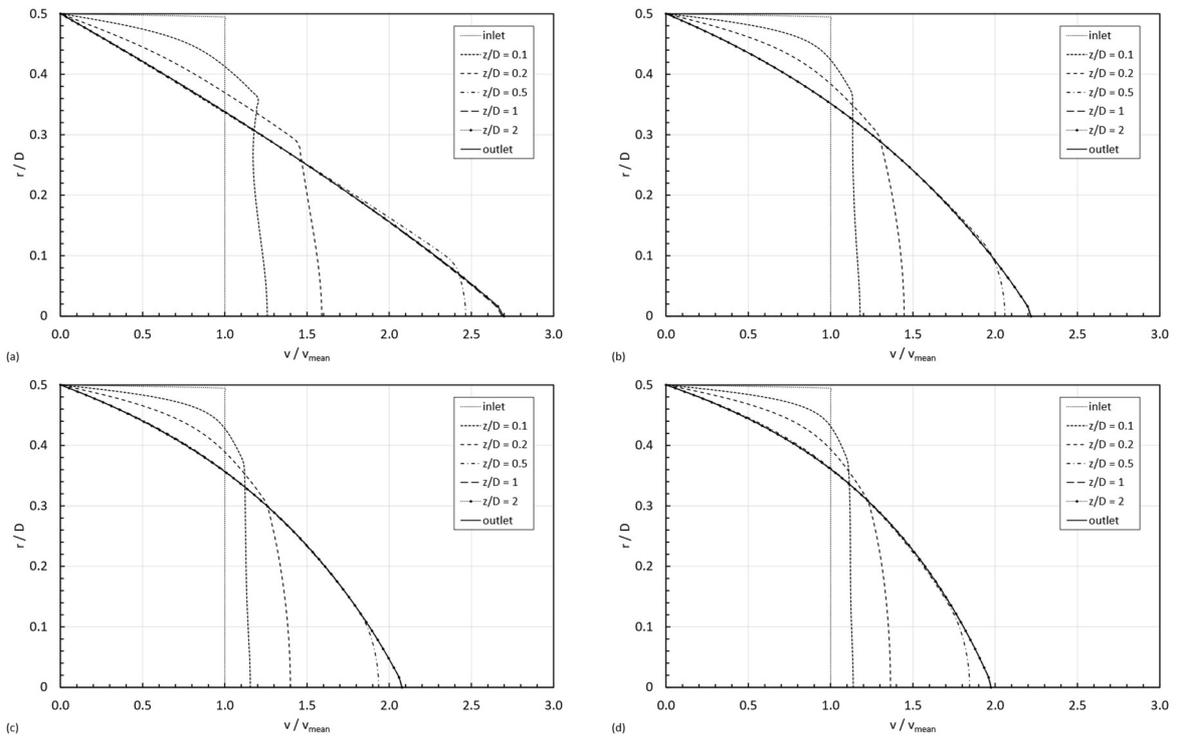

*Figure S3 – Velocity profiles for the STF 3 formulation in a D = 0.1mm pipe: $v_{in}$ = (a) 0.1mm/s, (b) 0.2mm/s, (c) 0.25mm/s and (d) 0.3mm/s.*

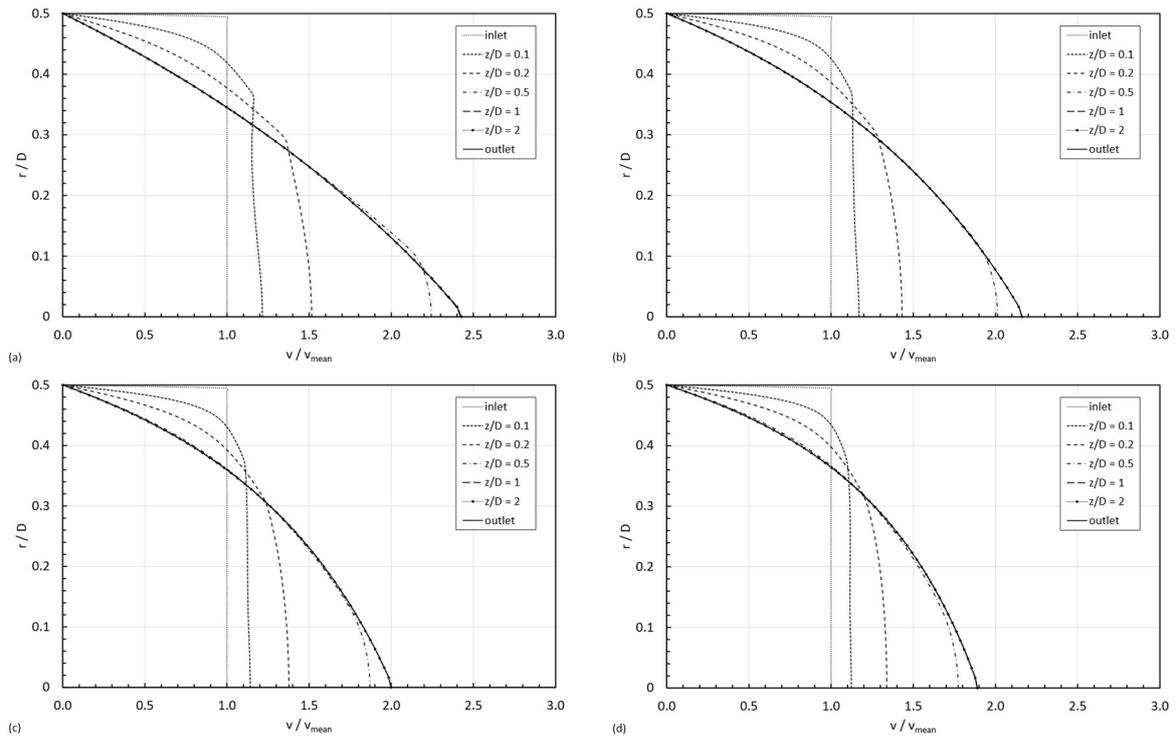

*Figure S4 – Velocity profiles for the STF 4 formulation in a D = 0.1mm pipe: $v_{in}$ = (a) 0.1mm/s, (b) 0.15mm/s, (c) 0.2mm/s and (d) 0.25mm/s.*